\documentclass[twocolumn,showpacs,preprintnumbers]{revtex4}
\usepackage{amsfonts}
\usepackage{amssymb}
\usepackage{amsmath}
\usepackage{graphicx}
\usepackage{dcolumn}
\usepackage{bm}

\begin{document}

\title{SU(1,1)-type light-atom correlated interferometer}
\author{Hongmei Ma$^{1}$, Dong Li$^{1}$, Chun-Hua Yuan$^{1,*}$, L. Q. Chen$%
^{1,\dag}$, Z. Y. Ou$^{1,2}$, and Weiping Zhang$^{1,\ddag}$}
\affiliation{$^{1}$Quantum Institute for Light and Atoms, Department of Physics, East
China Normal University, Shanghai 200062, P. R. China}
\affiliation{$^{2}$Department of Physics, Indiana University-Purdue University
Indianapolis, 402 North Blackford Street, Indianapolis, Indiana 46202, USA}

\begin{abstract}
The quantum correlation of light and atomic collective excitation can be
used to compose an SU(1,1)-type hybrid light-atom interferometer, where one
arm in the optical SU(1,1) interferometer is replaced by the atomic
collective excitation. The phase-sensing probes include not only the photon
field but also the atomic collective excitation inside the interferometer.
For a coherent squeezed state as the phase-sensing field, the phase
sensitivity can approach the Heisenberg limit under the optimal conditions.
We also study the effects of the loss of light field and the dephasing of
atomic excitation on the phase sensitivity. This kind of active SU(1,1)
interferometer can also be realized in other systems, such as circuit
quantum electrodynamics in microwave systems, which provides a different
method for basic measurement using the hybrid interferometers.
\end{abstract}

\maketitle

Enhanced phase estimation is important for high-precision measurements of
physical parameters \cite{Giovannetti11,Schnabel,Giovannetti06}. In optical
measurements, many physical parameters can be converted to phase
measurements of the optical field based on interferometer. The
phase-measurement precision can be described by quantum Fisher information
and the quantum Cram\'{e}r-Rao bound \cite{Helstrom67,Holevo82}. The
mean-square error in phase $\phi $ is given by the error-propagation
formula: $\delta \phi =\Delta A\left\vert d\langle A\rangle /d\phi
\right\vert ^{-1}$, where the average $\langle A\rangle $ and standard
deviation $\Delta A=\sqrt{\langle A^{2}\rangle -\langle A\rangle ^{2}}$ of
an observable $A$ are calculated in an optimal condition. High precisions
based on the interferometer can be reached by reducing the uncertainty $%
\Delta A$ or increasing the slope $d\langle A\rangle /d\phi $, or operating
them at the same time. The squeezed states \cite{Caves81,Xiao,Grangier,Pezze}
and two-mode squeezed states \cite{Ariano,Ole} are used to make the noise ($%
\Delta ^{2}A$) below vacuum noise. The slope can be improved by the
oscillation frequency or the amplitude enhancements of the output signal.
Beating the standard quantum limit (SQL) based on the oscillation frequency
enhancement was realized by the NOON states \cite{Dowling08,NOON}. The
amplitude improvement of the output signal was realized by changing the
structure of the interferometer, where the 50-50 beam splitters (BSs) in a
traditional Mach-Zehnder interferometer (MZI) were replaced by the optical
parameter amplifiers (OPAs) \cite{Yurke86}. This interferometer was
introduced by Yurke \emph{et al.} \cite{Yurke86} and is also called the
SU(1,1) interferometer because it is described by the SU(1,1) group, as
opposed to SU(2) for BSs. The SU(1,1) interferometry is under experimental
investigation by different groups \cite{Jing11,Lett} and even with
Bose-Einstein Condensates (BECs) \cite{Gross,Linnemann,Peise,Gabbrielli}.
Peise \emph{et al.} \cite{Peise} exploited the quantum Zeno effects using
the method of interaction-free measurements and Gabbrielli \emph{et al.}
\cite{Gabbrielli} realized a nonlinear three-mode interferometer, where the
analogy of optical down conversion, the basic ingredient of SU(1,1)
interferometry, is created with ultracold atoms.

\begin{figure}[ptb]
\centerline{\includegraphics[scale=0.42,angle=0]{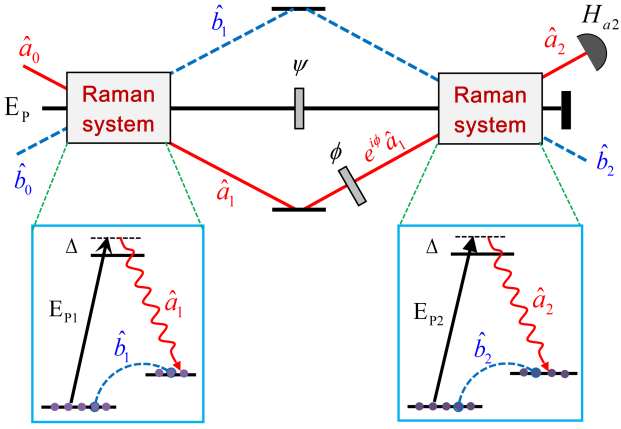}}
\caption{(Color online) (a) The schematic diagram of the SU(1,1)-type
light-atom correlated interferometer. In the optical SU(1,1) interferometer
of Yurke \emph{et al.}, two nonlinear beam splitters take the place of two
linear beam splitters in the traditional Mach-Zehnder interferometer (MZI).
Here, we use the Raman process (showed in the boxes) to produce a Stokes field $%
\hat{a}_{i}$ together with a correlated atomic collective excitation $\hat{b}%
_{i}$ ($i=1,2$) which are the beam splitting elements. That is, in the
light-atom-correlated interferometer, one arm is the Stokes field (solid
line) and the other arm is replaced by the atomic collective excitation (dashed
line). Two arms splitting and their recombination are composed of two Raman
processes which are successively implemented inside the same atomic system. $%
\hat{a}_{0}$ is the initial input light field. $\hat{b}_{0}$ is in vacuum or an
initial atomic collective excitation which can be prepared by another Raman
process or electromagnetically induced transparency process. The pump field $%
E_{p}$ between the two Raman processes has a $\protect\psi$ phase difference.
The output optical mode $\hat{a}_{2}$ is detected by the homodyne detector $%
H_{a2}$. $\protect\phi:$ phase shift.}
\label{fig1}
\end{figure}

Recently, an improved theoretical scheme was presented by Plick \emph{et al.}
\cite{Plick} who proposed to inject a strong coherent beam to
\textquotedblleft boost\textquotedblright\ the photon number. Experimental
realization of this SU(1,1) interferometer was reported recently \cite%
{Jing11}. The maximum output intensity of this interferometer can be much
higher than the input intensity as well as the intensity inside the
interferometer (the phase-sensing intensity). More recently, the noise
performance of this interferometer was analyzed \cite{Ou12,Li14}.
Experimentally, under the same phase-sensing intensity condition the
improvement of $4.1$ dB in signal-to-noise ratio of this interferometer over
a traditional linear interferometer was observed \cite{Hudelist}. Due to the
improved phase-measurement sensitivity of this interferometer, it was
suggested for gravitational wave detection, but it needs strong coherent
light input \cite{Plick}. The very strong coherent light will generate the
higher-order nonlinear effect and the radiation pressure noise. Combined
with the squeezed state input, the sensitivity of SU(1,1) can be improved
further due to the noise reduction \cite{Li14}. Collective atomic excitation
due to its potential applications for quantum information processing, has
attracted a great deal of interest \cite%
{Duan,Polzik,Kuzmich,Chen10,Jain,Yuan10,Yuan13,Chen15}. Here, we present an
SU(1,1)-type hybrid interferometer composed of the light and atomic
collective excitation. There are two advantages. One is high conversion
based on the Raman process. The other, more important one is that the phase
shift is from either or both the optical phase and the phase of the atomic
collective excitation which is sensitive to magnetic fields due to the
Zeeman effect. Such an interferometer should find wide applications in
precision measurement in atomic and optical physics. Our scheme presents an
extension and may be a substantial step forward in an SU(1,1) standard
optical interferometer.

In the SU(1,1)-type hybrid light-atom correlated interferometer discussed
here, we use a Raman process to produce a Stokes field together with a
correlated atomic collective excitation; that is, one arm in the optical
SU(1,1) interferometer is replaced by the atomic collective excitation. For
a coherent squeezed state as a phase-sensing field input, the phase
sensitivity can approach the Heisenberg limit. The effects of photon loss
and collisional dephasing of the atomic excitation on the the phase
sensitivity are analyzed.

Let us first introduce the SU(1,1)-type hybrid light-atom correlated
interferometer. In the optical SU(1,1) interferometer of Yurke \emph{et al.}
\cite{Yurke86}, two nonlinear beam splitters take the place of two linear
beam splitters in the traditional Mach-Zehnder interferometer (MZI). In the
SU(1,1)-type hybrid light-atom-correlated interferometer, one of two arms in
the optical SU(1,1) interferometer is replaced by an atomic collective
excitation as shown in Fig.~\ref{fig1}. In our scheme, the splitting and
recombination of the light field and atomic collective excitation are
composed of two Raman processes. In the first Raman process, similar to
beams splitting process, $\hat{b}_{0}$ is in vacuum, i.e., all atoms in
their ground hyperfine state $|1\rangle $ by optical pumping, or $\hat{b}_{0}
$ is an initial atomic collective excitation which can prepared by another
Raman process \cite{Chen10} or electromagnetically induced transparency
process \cite{Jain}. Then an off-resonant pump light $E_{P1}$ is applied to
the atomic ensemble together with a phase-sensing field $\hat{a}_{0}$,
generating stimulating Raman scattering $\hat{a}_{1}$ together with a
correlated atomic collective excitation $\hat{b}_{1}$ \cite{Chen10,Yuan10}.
At the same time, the incoming phase-sensing field $\hat{a}_{0}$ has been
amplified by the Raman process. In the second Raman process, similar to the
beams recombination process, the pump field $P_{2}$ together with the
generated Stokes field $\hat{a}_{1}$ inject the Raman system again. Before
the Stokes field $\hat{a}_{1}$\ injects the Raman system, it is subject to
phase $\phi $. After they inject the Raman system, the phase modulated
Stokes field $\hat{a}_{2}$ is generated as shown in Fig.~\ref{fig1} \cite%
{Yuan13,Chen15}. When the phase shift $\phi $ is $0$, the light field and
atomic collective excitation are decorrelated by the second Raman process.
But when the phase shift $\phi $ is not $0$, similar to the optical
interferometer, the phase-measurement sensitivity of this hybrid
interferometer is improved due to signal enhancement based on Raman
amplification, i.e., the slope of the output signal is improved. Compared
with the realization of an SU(1,1) interferometry via four-wave mixing \cite%
{Jing11,Hudelist}, the Raman process has high conversions due to the
second-order nonlinear process instead of the third-order nonlinear process.
In addition, the proposal can also introduce the atomic phase via a magnetic
field change into the phase measurement, and provide a different method for
basic measurements in atomic and optical physics.

\begin{figure}[ptb]
\centerline{\includegraphics[scale=0.45,angle=0]{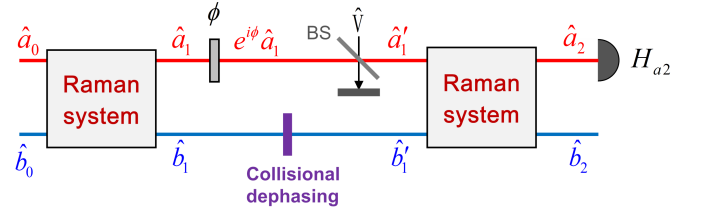}}
\caption{(Color online) A lossy interferometer model. The loss in the
optical arm is modeled by adding fictitious beam splitter, i.e., $\hat{a}%
^{\prime}_{1}=\protect\sqrt{T}\hat{a}_{1}(t_{1})e^{i\protect\phi}+\protect%
\sqrt{R}\hat{V}$ where $T$ and $R$ are the transmission and reflectance
coefficients, respectively, and $\hat{V}$ is in vacuum. The loss in the other arm is the
atomic collisional dephasing, i.e., $\hat {b}^{\prime}_{2}=\hat{b}%
_{1}(t_{1})e^{-\Gamma\protect\tau}+\hat{F}$ where $\Gamma$ is the collisional
dephasing rate, $\protect\tau$ is the delay between two Raman processes, and $%
\hat{F}$ is the noise operator. }
\label{fig2}
\end{figure}

Next, we analyze the two Raman processes. Considering a three-level
lambda-shaped atom system shown in the box of Fig. \ref{fig1}, the
Raman-scattering process is described by the following pair of coupled
equations \cite{Raymer04}:%
\begin{equation}
\frac{\partial \hat{a}(t)}{\partial t}=\eta A_{P}\hat{b}^{\dag }(t),\text{ \
}\frac{\partial \hat{b}(t)}{\partial t}=\eta A_{P}\hat{a}^{\dag }(t),
\label{eq1}
\end{equation}%
where $\hat{a}(t)$ is the light field operator, $\hat{b}(t)$ is the
collective atomic operator, $A_{P}$ is the amplitude of the pump field, and $%
\eta $ is the coupling constant. The solution of the above equation is $\hat{%
a}(t)=u(t)\hat{a}(0)+v(t)\hat{b}^{\dag }(0)$ and $\hat{b}(t)=u(t)\hat{b}%
(0)+v(t)\hat{a}^{\dag }(0)$, where $u(t)=\cosh (g)$, $v(t)=e^{i\theta }\sinh
(g)$, $g=\left\vert \eta A_{P}\right\vert t$, $e^{i\theta
}=(A_{P}/A_{P}^{\ast })^{1/2}$, and $t$ is the time duration of pump field $%
E_{P}$. Different initial conditions of $\hat{a}(0)$\ and $\hat{b}(0)$\
correspond to different scattering processes. We use different subscripts to
differentiate the two processes, where $1$ denotes the first Raman process
(RP1) and $2$ denotes the second Raman process (RP2). We first examine the
quantum correlations between the two modes $\hat{X}_{a}=(\hat{a}+\hat{a}%
^{\dagger })/2$\ and $\hat{X}_{b}=(\hat{b}+\hat{b}^{\dagger })/2$ of two
Raman processes. The correlation of light and atomic collective excitation
can be described by the linear correlation coefficient (LCC), which is
defined as \cite{Gerry}
\begin{equation}
J(\hat{X},\hat{Y})=\frac{cov(\hat{X},\hat{Y})}{\langle (\Delta \hat{X}%
)^{2}\rangle ^{1/2}\langle (\Delta \hat{Y})^{2}\rangle ^{1/2}},
\end{equation}%
where $\langle (\Delta \hat{X})^{2}\rangle =\langle \hat{X}^{2}\rangle
-\langle \hat{X}\rangle ^{2}$, $\langle (\Delta \hat{Y})^{2}\rangle =\langle
\hat{Y}^{2}\rangle -\langle \hat{Y}\rangle ^{2}$, and $cov(\hat{X},\hat{Y})=%
\frac{1}{2}(\langle \hat{X}\hat{Y}\rangle +\langle \hat{Y}\hat{X}\rangle
)-\langle \hat{X}\rangle \langle \hat{Y}\rangle $ is the covariance of the
quadrature phase operators $\hat{X}$ and $\hat{Y}$ \cite{Walls}. For RP1,
the LCC $J(\hat{X}_{a1},\hat{X}_{b1})$ is dependent on different initial
conditions. When $\hat{a}(0)$ and $\hat{b}(0)$ start from the vacuum, the
first RP1 created pair correlations. Here the pair is not a photon pair, but
is composed of a photon and an atomic collective excitation. The LCC is
given by%
\begin{equation}
J(\hat{X}_{a1},\hat{X}_{b1})=\cos \theta _{1}\tanh (2g_{1}).
\end{equation}%
For $\theta _{1}=0$ ($A_{P1}$ is real), the maximum LCC are $\tanh (2g_{1})$%
. The generated state is similar to the two-mode squeezed vacuum state $%
|\Psi \rangle =\sum_{n}c_{n}|n\rangle _{\text{ph}}|n\rangle _{\text{Atom}}$,
where $n$\ is the photon number and the number of atomic collective
excitation. If $\hat{a}(0)$ and $\hat{b}(0)$ are initially in coherent
states, the generated state is similar to a two-mode squeezed coherent state
\cite{Birrittella}. After RP1, the LCC is not zero, which shows that a
strong correlation exists between the number of photon and the number of
atomic collective excitation number.

\begin{figure}[ptb]
\centerline{\includegraphics[scale=0.45,angle=0]{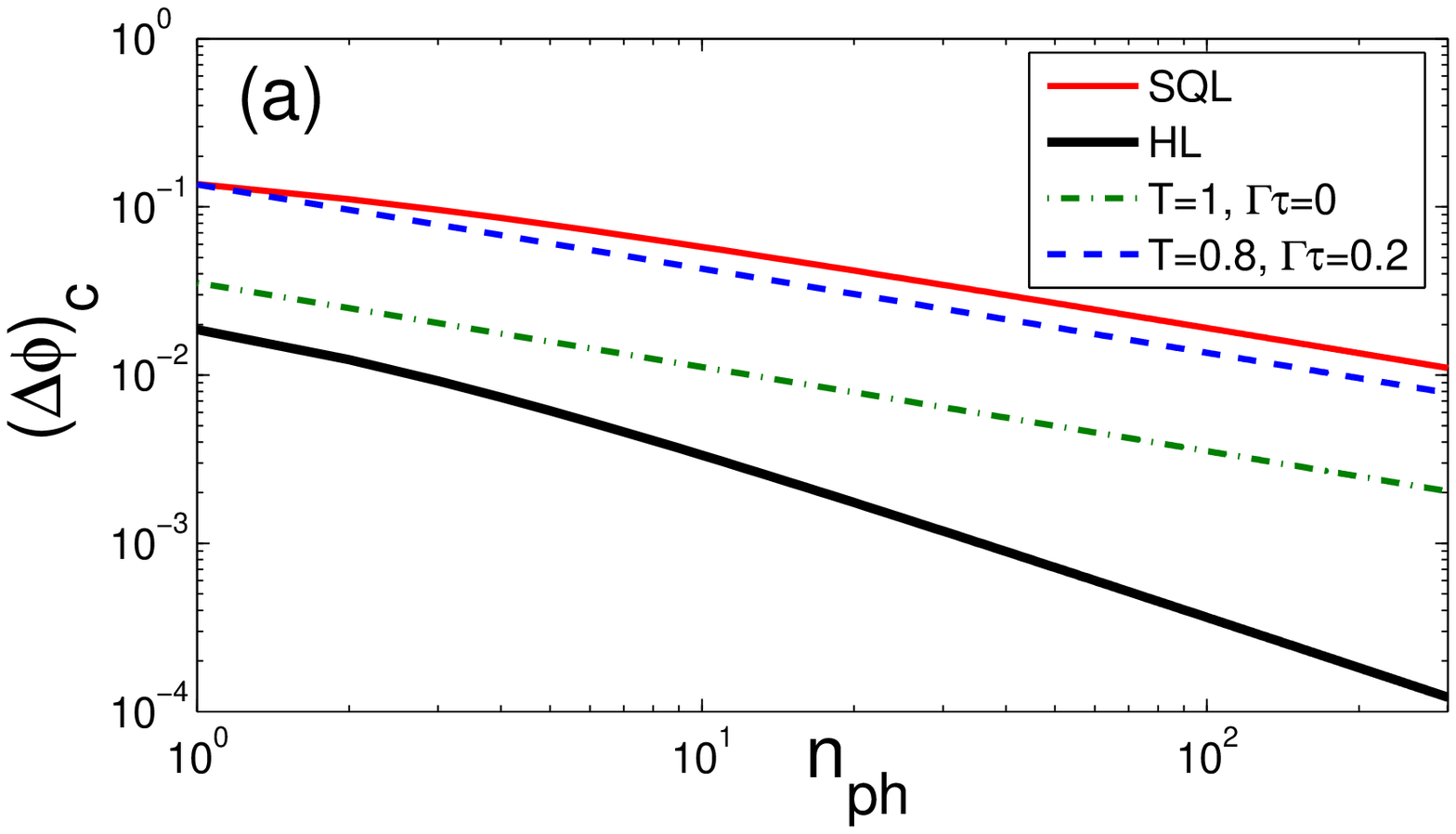}} %
\centerline{\includegraphics[scale=0.45,angle=0]{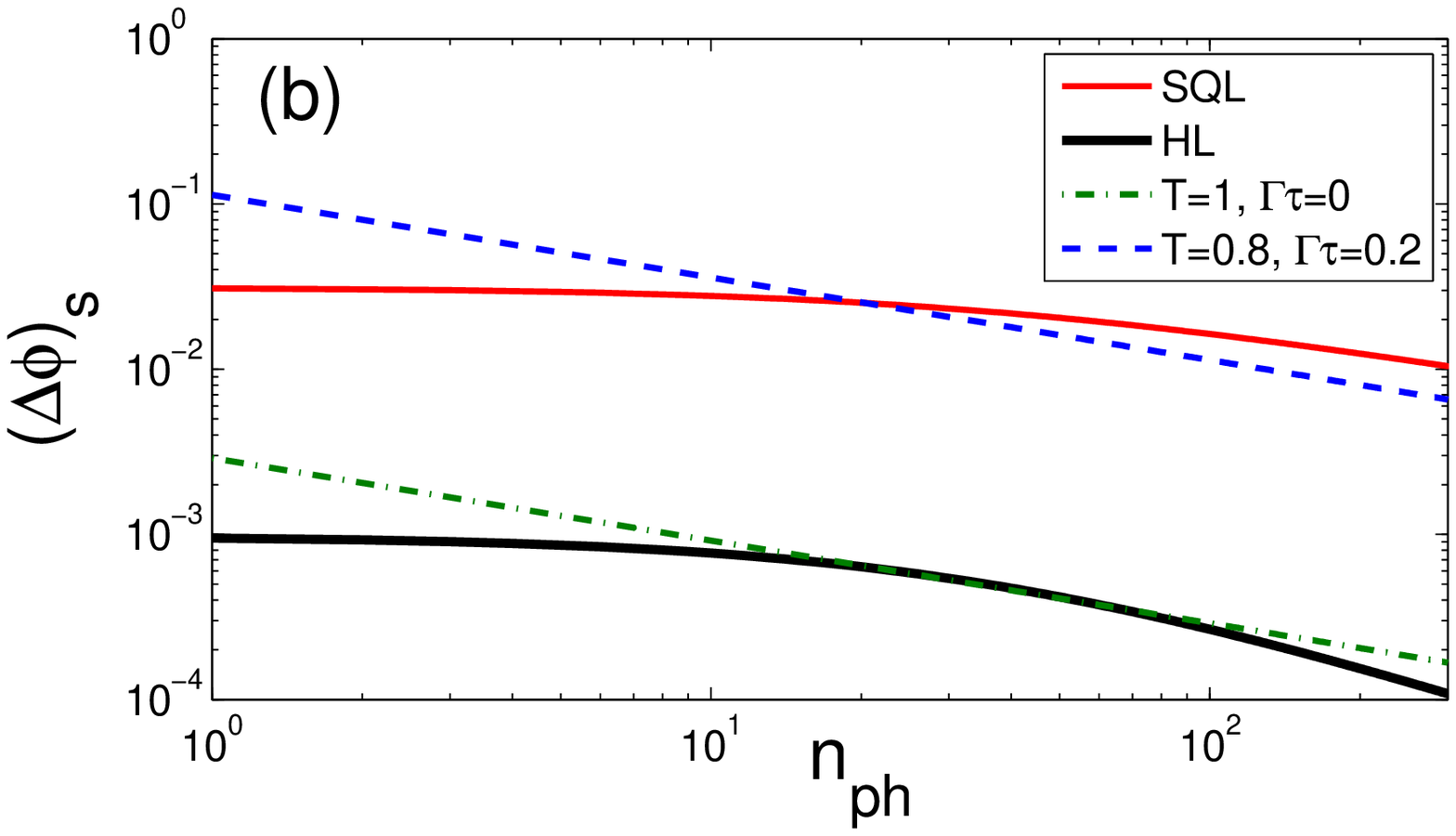}}
\caption{(Color online) The phase sensitivity (a) $\Delta\protect\phi_{c}$
and (b) $\Delta\protect\phi_{s}$ as a function of the number of probes $%
n_{ph}$ inside the interferometer with $g=1$. The input coherent squeezed
light with $r=2.5$}
\label{fig3}
\end{figure}

Then, we examine the quantum correlation of the RP2. After a delay time $%
\tau $ of the RP1 generation, the Stokes field $\hat{a}_{1}$ together with
the pumping field $E_{P2}$ drive the atomic system again shown in Fig.~\ref%
{fig1}. According to the solutions of Eq.~(\ref{eq1}), we can obtain $\hat{a}%
_{2}(t_{2})=u_{2}(t_{2})\hat{a}_{2}(0)+v_{2}(t_{2})\hat{b}_{2}^{\dag }(0)$
and $\hat{b}_{2}(t_{2})=u_{2}(t_{2})\hat{b}_{2}(0)+v_{2}(t_{2})\hat{a}%
_{2}^{\dag }(0)$, where $t_{2}$ is the duration of the pump field $E_{P2}$. $%
\hat{a}_{2}(0)$ and $\hat{b}_{2}(0)$\ are the initial conditions of RP2,
which is from the atomic collective excitation and stokes field of RP1. We
consider the collisional dephasing, which can be described by the factor $%
e^{-\Gamma \tau }$ (see Fig. \ref{fig2}). Then the initial condition $\hat{b}%
_{2}(0)$ is%
\begin{equation}
\hat{b}_{2}(0)=\hat{b}_{1}(t_{1})e^{-\Gamma \tau }+\hat{F},
\label{initial-b}
\end{equation}%
where $\tau $ is the delay and $\hat{F}=\int_{0}^{\tau }e^{-\Gamma (\tau
-t^{\prime })}f(t^{\prime })dt^{\prime }$. $\hat{f}(t)$ is the quantum
statistical Langevin operator describing the collision-induced fluctuation
and obeys $\langle \hat{f}(t)\hat{f}^{\dag }(t^{\prime })\rangle =2\Gamma
\delta (t-t^{\prime })$ and $\langle \hat{f}^{\dag }(t)\hat{f}(t^{\prime
})\rangle =0$. The Stokes light $\hat{a}_{1}$ is also subject to photon
loss, which can be equal to the effect of fictitious beam splitters inserted
in the channel, as shown in Fig.~\ref{fig2}. Considering the loss in the
propagation, the initial condition $\hat{a}_{2}(0)$ is given by
\begin{equation}
\hat{a}_{2}(0)=\sqrt{T}\hat{a}_{1}(t_{1})e^{i\phi }+\sqrt{R}\hat{V},
\label{initial-a}
\end{equation}%
where $T$ and $R$ are the transmission and reflectance coefficients,
respectively, and $\hat{V}$ is in vacuum.

Using the initial conditions given by Eqs.~(\ref{initial-b})-(\ref{initial-a}%
), the generated Stokes field $\hat{a}_{2}$ and atomic collective excitation
$\hat{b}_{2}$\ can be worked out:%
\begin{align}
\hat{a}_{2}(t_{2})& =\mathcal{U}_{1}\hat{a}_{1}(0)+\mathcal{V}_{1}\hat{b}%
_{1}^{\dag }(0)+\sqrt{R}u_{2}\hat{V}+v_{2}\hat{F}^{\dag }, \\
\hat{b}_{2}(t_{2})& =e^{-i\phi }[\mathcal{U}_{2}\hat{b}_{1}(0)+\mathcal{V}%
_{2}\hat{a}_{1}^{\dag }(0)]+\sqrt{R}v_{2}\hat{V}^{\dag }+u_{2}\hat{F},
\end{align}%
where $\mathcal{U}_{1}=\sqrt{T}u_{1}u_{2}e^{i\phi }+e^{-\Gamma \tau
}v_{1}^{\ast }v_{2}$, $\mathcal{V}_{1}=\sqrt{T}v_{1}u_{2}e^{i\phi
}+e^{-\Gamma \tau }u_{1}^{\ast }v_{2}$, $\mathcal{U}_{2}=e^{-\Gamma \tau
}u_{1}u_{2}e^{i\phi }+\sqrt{T}v_{1}^{\ast }v_{2}$, and $\mathcal{V}%
_{2}=e^{-\Gamma \tau }v_{1}u_{2}e^{i\phi }+\sqrt{T}u_{1}^{\ast }v_{2}$. When
$T=1$, $\Gamma \tau =0$, it reduced to the ideal lossless case, i.e., $%
\mathcal{U}_{1}=\mathcal{U}_{2}=\mathcal{U}$, $\mathcal{V}_{1}=\mathcal{V}%
_{2}=\mathcal{V}$, and $\mathcal{U}=[\cosh g_{1}\cosh g_{2}e^{i(\phi +\theta
_{1}-\theta _{2})}+\sinh g_{1}\sinh g_{2}]e^{i(\theta _{2}-\theta _{1})}$, $%
\mathcal{V}=[\sinh g_{1}\cosh g_{2}e^{i(\phi +\theta _{1}-\theta
_{2})}+\cosh g_{1}\sinh g_{2}]e^{i\theta _{2}}$, where $\left\vert \mathcal{U%
}\right\vert ^{2}-\left\vert \mathcal{V}\right\vert ^{2}=1$. When $\phi =0$
and $\theta _{2}-\theta _{1}=\pi $, we have $\mathcal{U}=1$ and $\mathcal{V}%
=0$. Therefore, under this condition, the LCC $J(\hat{X}_{a2},\hat{X}_{b2})$
is $0$ for any input states. That is, the RP2 will "undo" what the RP1 did.

Now, we analyze the phase sensitivity of this light-atom correlated
interferometer. In quantum phase precision measurement, the phase
sensitivity $\Delta \phi $ is defined by the linear error propagation%
\begin{equation}
(\Delta \phi )^{2}=\frac{\langle (\Delta \hat{O})^{2}\rangle }{\left\vert
\partial \langle \hat{O}\rangle /\partial \phi \right\vert ^{2}},
\label{sens}
\end{equation}%
where $\hat{O}$ is the measurable operator and $\langle (\Delta \hat{O}%
)^{2}\rangle =\langle \hat{O}^{2}\rangle -\langle \hat{O}\rangle ^{2}$. The
output amplitude quadrature operator $\hat{X}_{a2}=(\hat{a}_{2}+\hat{a}%
_{2}^{\dagger })/2$ is the detected variable. If we use the atomic variable $%
\hat{X}_{b2}=(\hat{b}_{2}+\hat{b}_{2}^{\dagger })/2$ as the detected
variable, we also get the phase sensitivity of the same order. But the
variable $\hat{X}_{b2}$ needs to read out by another light field. Here, we
consider the balanced situation is $g_{1}=g_{2}$, and $\theta _{2}-\theta
_{1}=\pi $.

For a coherent light $|\alpha \rangle $ ($|\alpha \rangle =\hat{D}(\alpha
)|0\rangle $, $\alpha =\left\vert \alpha \right\vert e^{i\theta _{\alpha }}$%
, $N_{\alpha }=\left\vert \alpha \right\vert ^{2}$) and a coherent squeezed
state $|\alpha ,\zeta \rangle $ ($|\alpha ,\zeta \rangle =\hat{D}(\alpha )%
\hat{S}(\zeta )|0\rangle $) as the phase-sensing fields, where $\hat{D}%
(\alpha )=e^{\alpha \hat{a}^{\dag }-\alpha ^{\ast }\hat{a}}$ and $\hat{S}%
(\zeta )=e^{(\zeta ^{\ast }\hat{a}^{2}-\zeta \hat{a}^{\dag 2})/2}$, $\zeta
=re^{i\theta _{s}}$, and the slopes of the output amplitude quadrature
operator $\hat{X}_{a2}$ are the same and are given by%
\begin{equation}
\left\vert \frac{\partial \langle \hat{X}_{a2}\rangle }{\partial \phi }%
\right\vert =\sqrt{TN_{\alpha }}\cosh ^{2}g\left\vert \sin (\phi +\theta
_{\alpha })\right\vert .  \label{slope}
\end{equation}%
When $\phi +\theta _{\alpha }=\pi /2$, the maximum slope is $\sqrt{%
TN_{\alpha }}\cosh ^{2}g$. But the output uncertainty is different, for the
coherent state and coherent squeezed-state input, and the uncertainties of
the output amplitude quadrature operator $\hat{X}_{a2}$ are given by $%
\langle (\Delta \hat{X}_{a2})^{2}\rangle _{c}=(\left\vert \mathcal{U}%
_{1}\right\vert ^{2}+\left\vert \mathcal{V}_{1}\right\vert
^{2})/4+[R\left\vert u_{2}\right\vert ^{2}+\left\vert v_{2}\right\vert
^{2}(1-e^{-2\Gamma \tau })]/4$, and $\langle (\Delta \hat{X}%
_{a2})^{2}\rangle _{s}=[\left\vert \mathcal{U}_{1}\right\vert
^{2}(e^{2r}\sin ^{2}\Theta +e^{-2r}\cos ^{2}\Theta )+\left\vert \mathcal{V}%
_{1}\right\vert ^{2}]/4+[R\left\vert u_{2}\right\vert ^{2}+\left\vert
v_{2}\right\vert ^{2}(1-e^{-2\Gamma \tau })]/4$, where $\Theta =\theta
_{s}/2+\theta _{\mathcal{U}_{1}}$ with $\mathcal{U}_{1}=\left\vert \mathcal{U%
}_{1}\right\vert e^{i\theta _{\mathcal{U}_{1}}}$, in which the second term $%
[R\left\vert u_{2}\right\vert ^{2}+\left\vert v_{2}\right\vert
^{2}(1-e^{-2\Gamma \tau })]/4$ is generated from the loss and collosional
dephasing.

\begin{figure}[ptb]
\centerline{\includegraphics[scale=0.45,angle=0]{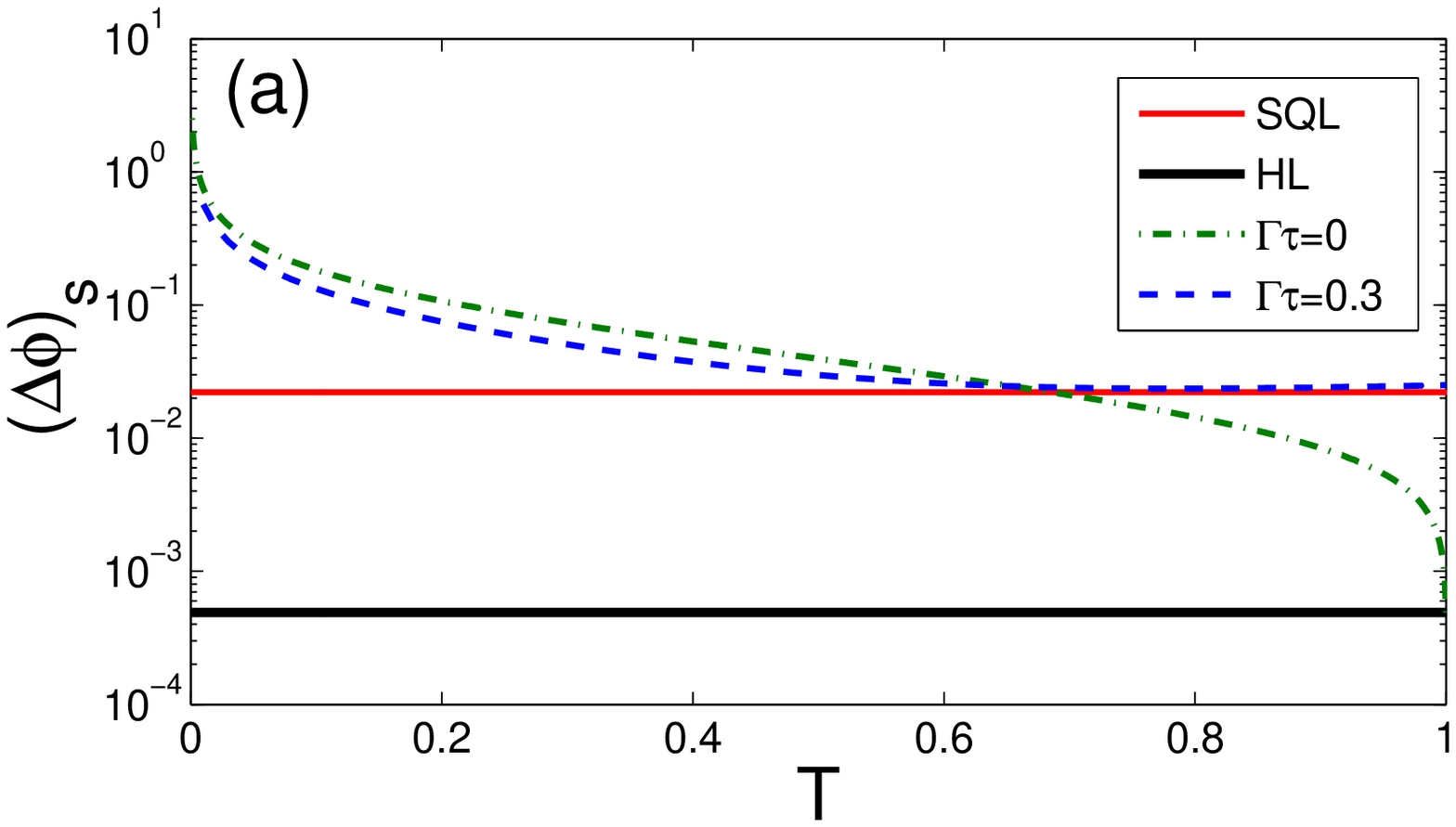}} %
\centerline{\includegraphics[scale=0.45,angle=0]{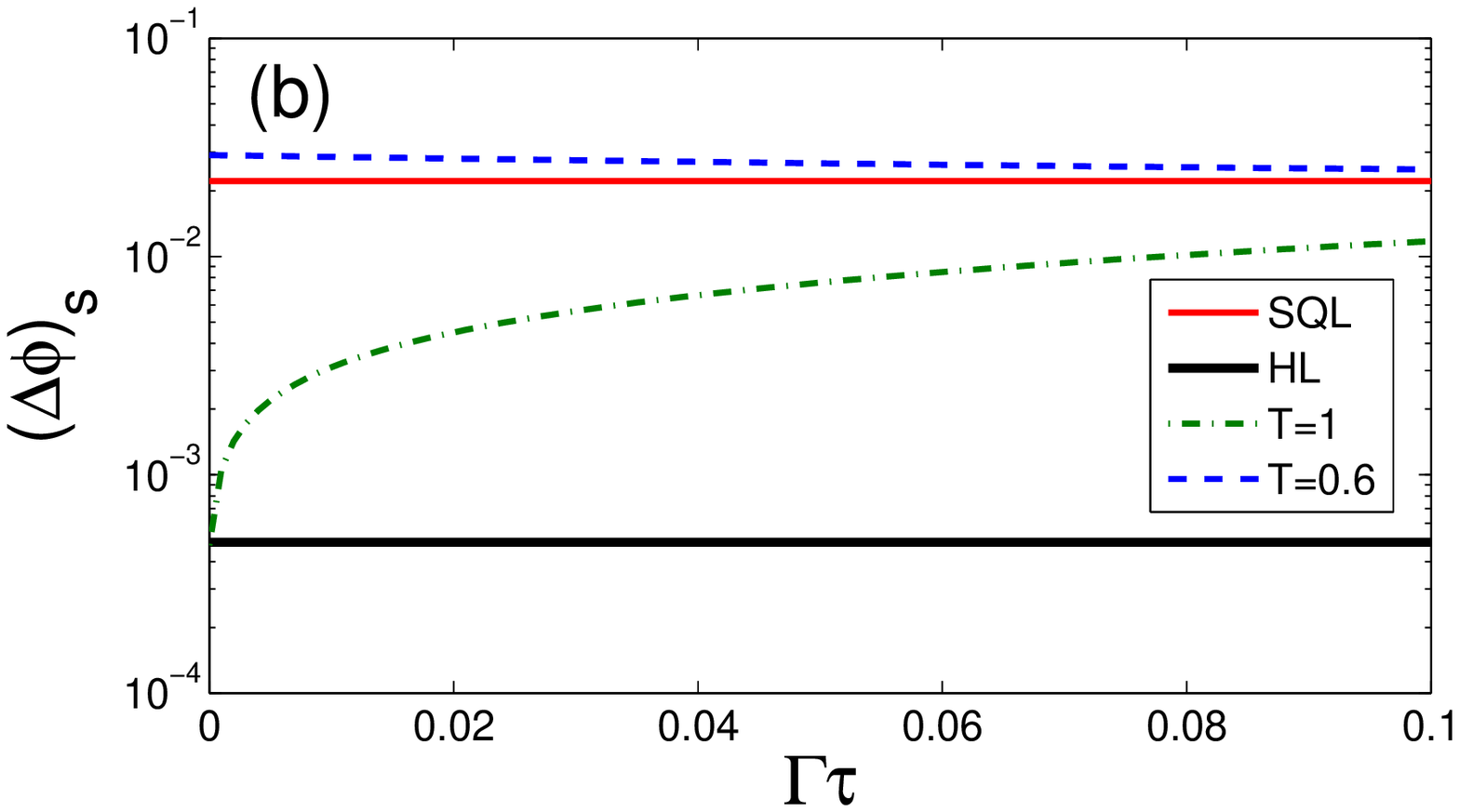}}
\caption{(Color online) The phase sensitivity $(\Delta\protect\phi)_{s}$ as
a function of (a) the transmission rate $T$ and (b) the collisional dephasing
rate $\Gamma\protect\tau$ with $r=2.5$, $N_{\protect\alpha}=e^{2r}/4$, $g=2$.
}
\label{fig4}
\end{figure}

The phase sensitivities $(\Delta \phi )_{c}$ and $(\Delta \phi )_{s}$ as a
function of the phase-sensing probe number $n_{\text{ph}}$ are shown in
Figs. \ref{fig3}(a) and \ref{fig3}(b), respectively. In the presence of the
loss and collisional dephasing, the phase sensitivities\ can beat the SQL
under the balanced situation. The phase sensitivity of the coherent
squeezed-state input is more easily affected by the loss and collisional
dephasing. Figs. \ref{fig4}(a) and \ref{fig4}(b) show the phase sensitivity $%
(\Delta \phi )_{s}$ as a function of the transmission rate $T$ and the
collisional dephasing rate $\Gamma \tau $, respectively. Under the condition
of $r=2.5$, $N_{\alpha }=e^{2r}/4$, $g=2$, when $\Gamma \tau \leqslant 0.3$
or $T\geq 0.6$, the phase sensitivity $(\Delta \phi )_{s}$ can beat the SQL.
Under the lossless ideal condition ($T=1$, $\Gamma \tau =0$) only $(\Delta
\phi )_{s}$ can reach the Heisenberg limit (HL). Now, we give the
explanations. Under the balanced and lossless situation, the uncertainty of
coherent state input is $\langle (\Delta \hat{X}_{a2})^{2}\rangle _{c}=1/4$,
which is from the vacuum fluctuation. For coherent squeezed state input, the
uncertainty is $\langle (\Delta \hat{X}_{a2})^{2}\rangle _{s}=e^{-2r}/4$
with $\Theta =0$, which is below the vacuum noise. The reduced noise can
improve the phase sensitivity. In the ideal case the phase sensitivities
under the optimal conditions are given by%
\begin{align}
(\Delta \phi )_{c}& =\frac{1}{\sqrt{N_{\alpha }}}\frac{1}{2\cosh ^{2}g},
\label{sens1} \\
(\Delta \phi )_{s}& =\frac{1}{e^{r}\sqrt{N_{\alpha }}}\frac{1}{2\cosh ^{2}g},
\label{sens2}
\end{align}%
which is improved by $\cosh ^{2}g$\ compared to the traditional Mach-Zehnder
interferometer for the same input phase-sensing field.

Now, we compare the optimal sensitivities given by Eqs. (\ref{sens1}) and (%
\ref{sens2})\ with SQL ($\varpropto 1/\sqrt{n_{\text{ph}}}$) and HL ($%
\varpropto 1/n_{\text{ph}}$). Here, the phase-sensing field is not the input
field as in the traditional MZI, but the amplified field inside the
interferometer. Although the phase shift is generated on the light field,
the light field and the atomic collective excitation are quantized. The
phase-sensing probe number includes not only the photon number $\langle \hat{%
a}_{1}^{\dag }(t_{1})\hat{a}_{1}(t_{1})\rangle $ but also the atomic
collective excitation number $\langle \hat{b}_{1}^{\dag }(t_{1})\hat{b}%
_{1}(t_{1})\rangle $, which is given by%
\begin{equation}
n_{\text{ph}}=N_{in}+N_{in}G_{RP}+G_{RP},  \label{Tot1}
\end{equation}%
where $G_{RP}=2\sinh ^{2}g$. The second term $N_{in}G_{RP}$ on the
right-hand side of Eq. (\ref{Tot1}) is the amplified signal of the input
photon number$\ $due to the stimulated Raman process, and the last term $%
G_{RP}$ is the number of amplified spontaneous-emission photons or noise.
For the coherent squeezed-state input case, the input photon number$\
N_{in}=\langle \alpha ,\zeta |\hat{a}_{1}^{\dag }(0)\hat{a}_{1}(0)|\alpha
,\zeta \rangle $ $=\left\vert \alpha \right\vert ^{2}+\sinh ^{2}r$. Under
the condition $G_{RP}\gg 1$, $\left\vert \alpha \right\vert \simeq
e^{r}/2\simeq \sinh r\gg 1$, the total phase-sensing probe number in the
interferometer can be written as $n_{\text{ph}}\simeq 2G_{RP}N_{\alpha }$.
The phase sensitivity of Eq. (\ref{sens2}) is given by%
\begin{equation}
(\Delta \phi )_{s}\simeq \frac{1}{2N_{\alpha }(G_{RP}+2)}\approx \frac{1}{n_{%
\text{ph}}}.  \label{HL}
\end{equation}%
From Eq. (\ref{HL}), with coherent squeezed state as input, the phase
sensitivity can reach HL due to the noise reduction and phase-sensing field
amplification. As has been previously pointed out, the loss is the limiting
factor in precision measurement \cite{Ou12,loss1,loss2,Dorner,Marino}. When
the transmission $T$ is close to $1$ and the collisional dephasing rate $%
\Gamma \tau $ is very small, the sensitivity is very high and can approach
the HL for the coherent squeezed-state input case, as shown in Fig. \ref%
{fig4}. Enhanced Raman scattering can be obtained by the initially prepared
atomic spin excitation \cite{Chen10,Yuan10} or by injecting a seeded light
field which is correlated with the initially prepared atomic spin excitation
\cite{Yuan13,Chen15}. This scheme is established on the basis of previous
studies and can be realized with high conversions. For Raman scattering, the
number of atoms must be bigger than the generated photon number, which is
easily realized for hot atoms.

In conclusion, the correlation between light and atomic collective
excitation can form an SU(1,1)-type hybrid light-atom-correlated
interferometer. The sensitivity is improved due to the signal enhancement
compared to the traditional MZI. When the transmission $T$ is close to $1$
and the collisional dephasing rate $\Gamma \tau $ is very small, the
sensitivity of the coherent squeezed-state input can approach the HL under
the optimal condition. This SU(1,1)-type hybrid light-atom-correlated
interferometer can generalize to other systems, such as circuit quantum
electrodynamics \cite{Barzanjeh}, which provides a different method for
basic measurement using the hybrid interferometers. The scheme can be
implemented with current technology.

This work is supported by the National Basic Research Program of China (973
Program) under Grant No. 2011CB921604, the National NSFC (Grants No.
11474095, No. 11274118, No. 11129402, and No. 11234003), and is supported by
the IPSMEC (Grant No. 13ZZ036) and the Fundamental Research Funds for the
Central Universities. \newline
Email: $^{\ast}$chyuan@phy.ecnu.edu.cn;\newline
$^{\dag }$lqchen@phy.ecnu.edu.cn; $^{\ddag}$ Team Leader.

\end{document}